\documentclass[reprint,aps,pra]{revtex4}
\usepackage{tabularx}
\usepackage{dcolumn}
\usepackage{graphics}
\usepackage{float}
\usepackage{bm}
\usepackage[dvips]{graphicx}

\usepackage{tabularx}
\usepackage{amsmath}
\usepackage{amssymb}
\usepackage{longtable}
\usepackage{verbatim}
\usepackage{float} 
\usepackage{fancyhdr} 
\usepackage{multirow}
\usepackage{mathtools,slashed}

\begin{document}
\pagestyle{fancy}
\rhead{\thepage}
\chead{}
\lhead{}
\cfoot{}
\rfoot{}
\lfoot{}

\title{Masses of Light Flavor Mesons using Bethe-Salpeter Approach}
\author{Iqra Liaqat}\affiliation{Centre For High Energy Physics, University of the Punjab, Lahore (54590), Pakistan. }
\author{Faisal Akram}\affiliation{Centre For High Energy Physics, University of the Punjab, Lahore (54590), Pakistan. }

\begin{abstract}
This work employs the approach based on the Bethe-Salpeter and Dyson-Schwinger equations to study the light meson spectrum. The Dyson-Schwinger equation of the quark propagator is truncated using the Maris-Tandy model for the dressed gluon propagator, which incorporates both the infrared enhancement and the perturbatively correct ultraviolet behavior. Additionally, for the dressed quark-gluon vertex, we apply its bare form and the Ball-Chiu model, which minimally satisfies the constraint imposed by the gauge symmetry through its Ward-Green-Takahashi identity. Consistent truncation of both Bethe-Salpeter and Dyson-Schwinger equations requires that axial-vector Ward-Takahashi identity, arising from chiral symmetry of the Lagrangian of quantum chromodynamics, must not be violated. We utilize this identity as an additional constraint to truncate the Bethe-Salpeter equation and examine the impact of two choices of quark-gluon vertex, along with the application of the Maris-Tandy model on the light meson spectrum. To extract the masses of flavored and unflavored light mesons, we solve the inhomogeneous Bethe-Salpeter equation using the Pad$\acute{e}$ approximation, which enables us to locate the poles of the Bethe-Salpeter amplitude without requiring a numerical solution in the timelike region of momentum space. We find that truncations of the Bethe-Salpeter equation based on bare and Ball-Chiu vertices, coupled with the Maris-Tandy model, yield consistent results.  
\end{abstract}
\maketitle

\section{Introduction}\label{intro}
Quantum chromodynamics (QCD) is the fundamental theory governing the strong interaction within the Standard Model of particle physics. It is a quantum field theory (QFT) in which fundamental fields are associated with quarks and gluons. However, due to the phenomenon of confinement, quarks and gluons are stuck within their bound states, known as hadrons. Despite the development of QCD approximately half a century ago, calculating the properties of hadrons from first-principle remains a formidable challenge within the realm of particle physics, primarily due to the fact that perturbative methods are invalidated at the energy scale at which quarks and gluons interact within the hadrons. Consequently, we find the use of several alternative approaches in the literature to calculate the properties of hadrons, including quark models \cite{GI81,GI82,GI83}, effective field theories \cite{GI95}, lattice QCD, and the continuum approach based on the combined use of the Dyson-Schwinger equations (DSEs) and Bethe-Salpeter equation (BSE) \cite{GI84,GI88,GI89,GI90,GI91}. DSEs, acting as the equations of motion of a QFT, are the exact coupled integral equations that the Green's functions of the QFT satisfy \cite{GI85,GI86,GI87}. However, BSE describes the bound states, such as mesons or baryons, in QCD. This approach allows us to consistently study non-perturbative strong interaction effects, including dynamical chiral symmetry breaking (DCSB), confinement, and bound states, within a single framework. DSEs relate dressed propagators and vertices, such as the dressed gluon propagator, quark propagator, and quark-gluon vertices, to each other, leading to infinitely coupled integral equations that connect various Green functions \cite{GI92,GI93,GI94,GI95,GI96,GI97,GI98,GI99}. Numerical studies, however, require implementing some truncation scheme to decouple equations of dressed propagators, based on modeling higher Green functions, such as the quark-gluon vertex (QGV).
Any acceptable Ans$\ddot{a}$tz of QGV \cite{GI100,GI101,GI102,GI103,GI104,GI105,GI106} must satisfy general field theoretical constraints, including: 1) the Ward-Green-Takahashi identity (WGTI), ensuring gauge covariance and invariance \cite{GI112}, 2) be free from kinematic singularities \cite{GI111}. 3) supporting the multiplicative renormalizability of the dressed propagators \cite{GI107,GI108}, and 4) agreeing with perturbation theory in the weak coupling limit \cite{GI109,GI110}.
It turns out that the chiral symmetry of the QCD Lagrangian relates pseudo-scalar and axial-vector Bethe-Salpeter amplitudes (the solutions of corresponding BSEs) through an identity, called the axial-vector Ward-Takahashi identity (avWTI) \cite{GI125}, which also involves dressed-quark propagator.
 This implies that in any reliable study of light-quark hadrons, a truncation scheme implemented on DSEs must also be consistently implemented on BSE so that the avWTI may not be violated. It is found that the rainbow-ladder (RL) truncation scheme \cite{GI113,GI114}, based on replacing the dressed QGV by the bare vertex, does not break avWTI when consistently implemented on the gap equation (DSE of the quark propagator) and BSE. 
 A limitation of the RL truncation is that it does not satisfy the full Slavnov-Taylor identity (STI) for the QGV, which involves an unknown quark-ghost kernel. A commonly used improvement is the Ball-Chiu (BC) vertex \cite{GI102}, which satisfies the Abelian WTI for a trivial ghost-gluon kernel, while its longitudinal part is constrained and the unknown transverse part is set to zero. More importantly, the BC vertex also preserves the axWTI, ensuring the correct realization of chiral symmetry. In this case, avWTI requires truncating the BSE in a way different from the RL truncation. This improvement has been studied in Ref. \cite{GI125} where the ultraviolet-finite model gap equation is used both within and beyond the RL truncation, using the BC vertex to calculate the pion and sigma meson spectra. In our calculations, we improve the ultraviolet-finite model by employing the full form of the MT model \cite{GI115,GI118}, which includes both the infrared enhancement and the ultraviolet behavior, using the DSE-BSE approach. We also extend the study to flavored mesons, such as kaons, both within and beyond the RL truncation, as well as to vector mesons within the RL truncation. 


   The paper is structured in the following form:\\
In Sec. 2, the DSE formalism is introduced, including details on the truncation schemes and the Ans$\ddot{a}$tze \cite{GI115,GI125} we are considering for specifying the kernel for the quark DSE. Introducing the MT model  \cite{GI109} and presenting the essential components for calculation. 
Additionally, we provide a detailed discussion of the vertex constructions employed, focusing on the Ans$\ddot{a}$tze used in recent SDE research employing the gauge approach. The algebraic equations for the gap equation's kernels resulting from vertex selection are discussed.
In Sec. 3, we introduce the bound state BSE. To analyze and solve the BSE, we used Dirac covariants designed to satisfy CPT constraints and achieve trace orthogonality \cite{GI117}. 
The BS amplitude is defined and expressed in the form of invariant amplitudes, and the method for solving the (in)homogeneous BSE is also explained.
  In Sec. 4,  numerical results are presented. Finally, a summary and conclusion are provided in Sec. 5, and we also discuss possible future extensions of this work.

\section{QUARK DYSON-SCHWINGER EQUATION}\label{conv.model}
\begin{figure}[H]
  \centering
  \includegraphics[width=0.7\linewidth]{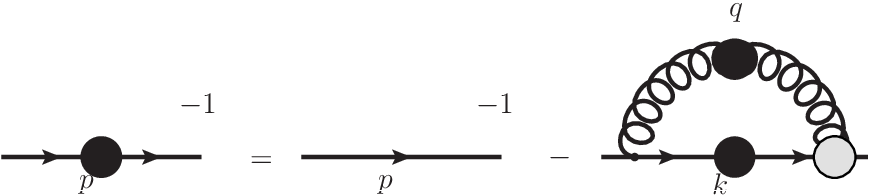}
 \caption{The Quark DSE; filled circles represent full propagators.}\label{sde}
\end{figure}

SDEs are a helpful non-perturbative tool for exploring confinement and DCSB \cite{GI128,GI129,GI130,GI150,GI151,GI152,GI160,GI162}. The renormalized form of this equation for the dressed-quark propagator is:
\begin{eqnarray}\label{Q.P}
S^{-1}(p)=Z_{2}(i\slashed{p}+m_{0})+\Sigma^{0}(p)
\end{eqnarray}
The self-energy of the quark, $\Sigma^{0}(p)$, is defined as follows:
\begin{eqnarray}\label{4}
\Sigma^{0}(p)=Z_{1}g^{2}\int_{k}^{\Lambda}D_{\mu\nu}(q)\frac{\lambda^{c}}{2}\gamma_{\mu}S(k)\Gamma_{\nu}^{c}(k,p)
\end{eqnarray}
Where, $q=k-p$ and $\int_{k}^{\Lambda}=\int^{\Lambda}\frac{d^4k}{(2\pi)^{4}}$, and $m_{0}$ represents the bare quark mass, $Z_{1}$ and $Z_{2}$ are the renormalization constants for the QGV and the quark wave-function, respectively, both depend on the renormalization point $(\mu)$ and the regularization mass scale $(\Lambda)$.
The renormalized dressed-gluon propagator is denoted by $D_{\mu\nu}(q)$, the dressed-quark propagator by $S(k)$, and the renormalized dressed QGV by $\Gamma_{\nu}^{c}(k,p)$. See Fig. \ref{sde}.
The general structure of $S(p)$ is written as:
\begin{eqnarray}\label{s.eq}
 S^{-1}(p)=i\slashed{p}A(p^{2})+B(p^{2})
\end{eqnarray}
Where, $A(p^{2})$ and $B(p^{2})$ represent the scalar and vector amplitudes.
At the renormalization point
\begin{eqnarray}\label{con.potential}
 S^{-1}(p)|_{p^{2}=\mu^{2}}=i\slashed{p}+m(\mu)
\end{eqnarray}
Where, $m(\mu)$ represents the current quark mass.
 The gluon propagator can be expressed as:
\begin{eqnarray}\label{con.potential}
D_{\mu\nu}(q)=\frac{D(q^{2})}{q^{2}}[\delta_{\mu\nu}-\frac{q_{\mu}q_{\nu}}{q^{2}}]+\xi\frac{q_{\mu}q_{\nu}}{q^{4}}
\end{eqnarray}
where $D(q^{2})$ is related to vacuum polarization by $D(q^{2})=\frac{1}{1+\Pi(q^{2})}$ and $\xi$ denotes the covariant gauge parameter with $\xi=0$ indicating the Landau gauge.
Using Ans$\ddot{a}$tz \cite{GI115},
\begin{eqnarray}\label{gluon.prop}
Z_{1}g^{2}D_{\mu\nu}(q)\frac{\lambda^{c}}{2}\gamma_{\mu}S(k)\Gamma_{\nu}^{c}(k,p) \rightarrow \frac{g(q^{2})}{q^{2}}\hat{\Delta}_{\mu\nu}(q)\frac{\lambda^{c}}{2}\gamma_{\mu}S(k)\Gamma_{\nu}^{c}(k,p)
\end{eqnarray}
Where, $g$ is the coupling constant,  $\hat{\Delta}_{\mu\nu}(q)= \delta_{\mu\nu}-\frac{q_{\mu}q_{\nu}}{q^{2}}$ is the transverse operator in the Landau gauge and $\frac{g(q^{2})}{q^{2}}$, as suggested by the MT model, is expressed as \cite{GI118,GI131,GI138}.
\begin{eqnarray}\label{mt.model}
\frac{g(q^{2})}{q^{2}}=\frac{4\pi^{2}}{\omega^{6}}Dq^{2}e^{-\frac{q^{2}}{\omega^{2}}}+4\pi\frac{\gamma_{m}\pi}{\frac{1}{2}ln(\tau+(1+\frac{q^{2}}{\Lambda_{QCD}^{2}})^{2})}f(q^{2})
\end{eqnarray}
With $f(q^{2}) = \frac{1-e^{-\frac{q^{2}}{4m_{t}^{2}}}}{q^{2}},$ $\tau=e^2-1,\;$$\gamma_{m}=\frac{12}{33-2N_{f}}$$,\;\Lambda_{QCD}=0.234\;$GeV, $D=0.93$\;GeV$^{2},\;$ $\omega=0.4 \;$GeV,$\;m_{t}=0.5\;$GeV$,\;N_{f}=4$,$\;m_{u/d}=0.000374$$\;$GeV. The parameter $m_{t}$ sets the scale between the perturbative and nonperturbative regions of Eq. \ref{mt.model}. $\omega$ and $D$ represent variable parameters in the model. According to \cite{GI153}, the calculated chiral condensate, as well as the masses and decay constants for the $\pi$ and $\rho$ mesons, vary minimally when $D$ is varied, provided that the product remains constant, with $\omega$ in the range of 0.3-0.5 GeV. Consequently, a configuration where $D.\omega$ remains constant can be seen as a single-parameter model, primarily defined by the need to achieve the correct values for $m_{\pi}$ and $f_{\pi}$ \cite{GI127,GI132,GI133,GI134}.
\subsection{DSE with bare vertex}\label{conv.model}
To solve the DSEs, truncations are necessary, and a recent analysis may be found in Refs. \cite{GI95,GI128,GI90,GI135,GI136,GI137,GI115}.
The RL truncation is the approximation, 
in which the dressed QGV $\Gamma_{\nu}(k,p)$ is replaced by the bare vertex $\gamma_{\nu}$. In Eq. \ref{gluon.prop}, the quark self-energy is expressed as:
\begin{equation}
\begin{aligned}
\Sigma^{0}(p)=Z_{1}g^{2}\int_{k}^{\Lambda}D_{\mu\nu}(q)\frac{\lambda^{c}}{2}\gamma_{\mu}S(k)\frac{\lambda^{c}}{2}\gamma_{\nu}
\end{aligned}
\end{equation}
By substituting the general form given in Eq. \ref{s.eq} and performing suitable Dirac traces on Eq. \ref{Q.P}, we can extract the Dirac-odd and 
Dirac-even components, which are written as follows:
 \begin{equation}\label{dres.fun}
\begin{aligned}
B\left(p^2\right)= & m_0 Z_2+4 \int_k \frac{g\left(q^2\right)}{q^2} \sigma^s\left(k^2\right), \\
A\left(p^2\right)= & Z_2+\frac{4}{3 p^2} \int_k \frac{g\left(q^2\right)}{q^2}\sigma^v\left(k^2\right)\left[k \cdot p+\frac{2 k \cdot q p \cdot q}{q^2}\right].
\end{aligned}
\end{equation}
Where, \\
\begin{equation}\label{dresss.fun}
\begin{aligned}
\sigma^s\left(k^2\right)= \frac{B\left(k^2\right)}{k^2 A^2\left(k^2\right)+B^2\left(k^2\right)}, \\
\sigma^v\left(k^2\right)= \frac{A\left(k^2\right)}{k^2 A^2\left(k^2\right)+B^2\left(k^2\right)}.
\end{aligned}
\end{equation}

\subsection{The dressed-quark-gluon vertex}\label{rel wave eq}


The non-perturbative fermion-boson vertex, $\Gamma_{\mu}(p,k)$, must satisfy several constraints \cite{GI104}. Not only does it obey its own DSE, but it must also satisfy the WGTI \cite{GI111,GI139,GI140}.\\
 In the limit $k \rightarrow p$, this reduces to WI:
 
\begin{equation}\label{relp.eq}
\Gamma_{\mu}(p,p)=\frac{\partial S^{-1}(p)}{\partial p_{\mu}}
\end{equation}
We can write the vertex in the form of transverse ($\Gamma_{\mu}^{\hat{T}}$) and longitudinal $(\Gamma_{\mu}^{\hat{L}})$ components.
\begin{equation}\label{rel.eq}
\Gamma_{\mu}(k,p)=\Gamma_{\mu}^{\hat{L}}(k,p)+\Gamma_{\mu}^{\hat{T}}(k,p)
\end{equation}
WGTI is given as:
\begin{equation}\label{rel.eq}
iq_{\mu}\Gamma^{\mu}(k,p)=S^{-1}(k)-S^{-1}(p)
\end{equation}
The $\Gamma_{\mu}^{\hat{T}}$ satisfies $q_{\mu}\Gamma_{\hat{T}}^{\mu}(k,p)=0$, which implies that the transverse part is unaffected by the WGTI, leaving only the longitudinal part to satisfy Eq. \ref{rel.eq}:
\begin{equation}\label{rell.eq}
iq_{\mu}\Gamma^{\mu}(k,p)=iq_{\mu}\Gamma_{\hat{L}}^{\mu}(k,p)=S^{-1}(k)-S^{-1}(p)
\end{equation}
Using the inverse fermion propagator in Eq. \ref{relp.eq}, the longitudinal component of the vertex, known as BC vertex, is given by:
\begin{equation}\label{BCL}
\begin{aligned}
i\Gamma_\mu(k, p)=i\Gamma_\mu^{\hat{L}(B C)}(k, p)= i\bar{\Delta}_A(k^2, p^2)\gamma_\mu+2\ell_{\mu}[i\gamma.\ell\Delta_A(k^2, p^2)+\Delta_B(k^2, p^2)],
\end{aligned}
\end{equation}
Given that $\ell=(p+k)/2$,\;\;$\bar{\Delta}_{\phi_{0}}(k^2, p^2)=\frac{[\phi_{0}(k^2)+\phi_{0}(p^2)]}{2}$,\;\;$\Delta_{\phi_{0}}(k^2, p^2)=\frac{[\phi_{0}(k^2)-\phi_{0}(p^2)]}{k^2- p^2}$
\cite{GI141,GI110,GI142}.\\
{\fontsize{10.5}{16}\selectfont \textbf{DSE with Ball-Chu vertex}}\\
We now use the BC vertex to derive the DSE equation describing the quark propagator. Returning to Eq. \ref{Q.P}, if the BC vertex is used, we find that:
 \begin{equation}\label{dres.fun}
\begin{aligned}
B\left(p^2\right)= & m_0 Z_2+\frac{4}{3}  \int_k \frac{g\left(q^2\right)}{q^2}
& \times\left\{\sigma^s\left(k^2\right)\left[L_{B 1}+L_{B 2}\right]-\sigma^v\left(k^2\right)\left[L_{B 3}\right]\right\},
\end{aligned}
\end{equation}
where $L_{B 1}, L_{B 2}$ and $L_{B 3}$ are associated with the BC vertex, given that:
\begin{equation}\label{dres.fun}
\begin{aligned}
& L_{B 1}=3\bar{\Delta}_A\left(k^2, p^2\right), \\
& L_{B 2}=\Delta_A\left(k^2, p^2\right)\left\{\frac{w^2 q^2-(w \cdot q)^2}{2 q^2}\right\}, \\
& L_{B 3}=\Delta_B\left(k^2, p^2\right)\left\{\frac{q^2 w \cdot k-w \cdot q q \cdot k}{q^2}\right\} .
\end{aligned}
\end{equation}
\begin{equation}\label{dres.fun}
\begin{aligned}
A\left(p^2\right)= & Z_2+\frac{4}{3}  \int_k \frac{g\left(q^2\right)}{q^2}
& \times\left\{\sigma^v\left(k^2\right)\left[L_{A 1}-L_{A 2}\right]+\sigma^s\left(k^2\right)\left[L_{A 3}\right]\right\},
\end{aligned}
\end{equation}
where $L_{A 1}, L_{A 2}$ and $L_{A 3}$ are written as:
\begin{equation}
\begin{aligned}
L_{A 1}= & \frac{\bar{\Delta}_A\left(k^2, p^2\right)}{p^2} 
\left\{\frac{k \cdot p q^2+2\left[\left(k^2+p^2\right) k \cdot p-k^2 p^2-(k \cdot p)^2\right]}{q^2}\right\}, \\
L_{A 2}= & \frac{\Delta_A\left(k^2, p^2\right)}{2 p^2} 
\left\{\left[p^2 k+k^2 p\right] \cdot w-\frac{p^2 w \cdot q k \cdot q-k^2 w \cdot q p \cdot q}{q^2}\right\}, \\
L_{A 3}= & \Delta_B\left(k^2, p^2\right) \frac{1}{p^2}\left\{\frac{w \cdot q p \cdot q-w \cdot p q^2}{q^2}\right\}.
\end{aligned}
\end{equation}
Where, $w=k+p$.
\section{THE BETHE-SALPETER EQUATION }
\begin{figure}[H]
   \centering
  \includegraphics[width=0.8\linewidth]{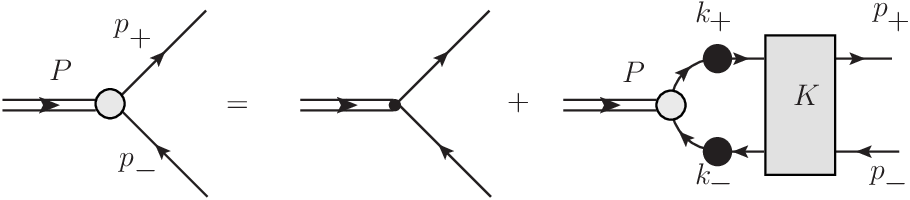}
  \caption{In Homogeneous BSE}\label{ibse}
\end{figure}

This section covers the fundamental study of the (in)homogeneous BSE in the context of meson bound states.\\
The BSE is the most established method for addressing the relativistic two-body system in QFT \cite{GI119,GI156}, and is used to describe $q\bar{q}$ bound states \cite{GI120,GI121,GI158,GI161}.
We will focus on equations describing the bound states of mesons, with $\Gamma_{M}^{fg}$ representing a quark-meson vertex \cite{GI115,GI122,GI157,GI144}. The renormalized inhomogeneous BSE for the $ q\bar{q}$ system is stated as
\begin{eqnarray}\label{homo.eq}
[\Gamma_{M}^{fg}(p,P)]_{tu}=[\Gamma_{0}^{fg}(p, P)]_{tu} +\int\frac{dk^{4}}{(2\pi)^{4}}K_{tu}^{rs}(p,k;P)[X_{M}^{fg}(k,P)]_{rs}
\end{eqnarray}
\begin{eqnarray*}\label{con.potential}
[X_{M}^{fg}(k,P)]_{rs}=[S^f(k_{+})\Gamma_{M}^{fg}(k;P)S^g (k_{-})]_{rs}
\end{eqnarray*}
 $[X_{M}^{fg}(k,P)]_{rs}$ represents the Bethe-Salpeter wave function. Here, $P^{2}=-m^{2}$, $m$ represents the bound state mass, $p$ represents the $q\bar{q}$ momentum, and $r, s, t,$ and $u$ denote the color, and Dirac indices. Kernel $K$ denotes the scattering kernel for the $q\bar{q}$ interaction. Here, $k_{\pm}=k\pm \eta_{\pm}P$, and we choose $\eta=1/2$. $P$ represents the total momentum of the meson. See Fig. \ref{ibse}.
  
For S and PS mesons, $M$ is simply specifying the type of meson, for V and AV mesons, the Eq. \ref{homo.eq} has a Lorentz index, further expanding the complexity of the system Ref. \cite{GI115}.
\subsection{BSE with bare vertex}\label{conv.model}
\begin{figure}[H]
   \centering
  \includegraphics[width=0.4\linewidth]{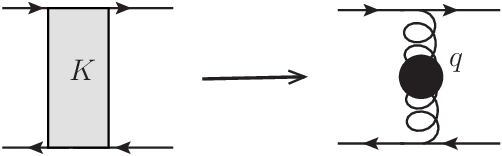}
  \caption{Scattering kernel with ladder approximation}\label{bbbb}
\end{figure}

Starting from Eq. \ref{homo.eq}, we implement the ladder truncation by using the bare vertex; see Fig. \ref{bbbb}.
\begin{equation}\label{gama.fun}
\begin{aligned}
K_{tu}^{rs}(p,k;P)\rightarrow -g(q^2)D_{\mu\nu}^{0}(q)[\frac{\lambda^c}{2}\gamma_{\mu}]^{ru}\otimes[\frac{\lambda^c}{2} \gamma_{\nu}]^{ts}
\end{aligned}
\end{equation}
where $K_{tu}^{rs}(p,k;P)$ represents the renormalized $q\bar{q}$ scattering kernel, and mesons with different quantum numbers exhibit different forms of $\Gamma_{M}^{fg}(p;P)$. Generally, $\Gamma_{M}^{fg}(p;P)$ is generated by considering all possible combinations of $p, P$  and $\lambda$ matrices.\\ 
The BSA, $\Gamma_{M}^{fg}(p;P)$, depends on the total momentum $P$ and the relative momentum $p$ of the $q\bar{q}$ pair. Using Lorentz-invariant variables, the amplitudes can be expressed as a function of $P^{2},\;\;p^{2},$ and $p.P$. 
A suitable basis for the amplitudes can be constructed from the product of Dirac matrices. The general form of the BSA can then be written as a linear combination of $N$ covariant entities denoted by $T^{i}$, each multiplied by a corresponding scalar amplitude represented as $F_{i}$, where $i$ ranges from $1$ to $N$.
\begin{equation}\label{gama1.fun}
\begin{aligned}
\Gamma^{fg}_{M}(p;P)=\sum^{N}_{i=1}F_{i}(p^{2},P^{2},p.P)T^{i}_{M}(p;P)
\end{aligned}
\end{equation}
The BSE for a PS meson with $J^{PC}=0^{-+}$ can be written as follows:
\begin{equation}\label{gama.fun}
\begin{aligned}
i \Gamma^{uu}(p;P)=i\gamma_{5}E_{\pi}(p;P)+\gamma_{5}\gamma.PF_{\pi}(p;P)+\gamma_{5}\gamma.pG_{\pi}(p;P)+\gamma_{5}p_{\alpha}\sigma^{\alpha\beta}P_{\beta}H_{\pi}(p;P)
\end{aligned}
\end{equation}
Since $\pi^{0}$ is a $C=+1$ charge eigenstate that obeys $C|\pi^{0}\rangle = +|\pi^{0}\rangle$, the BSA satisfies the condition as mentioned in Ref. \cite{GI109}, namely, $\bar{\Gamma}^{uu}(p;P)=\Gamma^{uu}(p;P)$.
If an orthonormal basis of covariants is chosen such that $Tr_{D}[T^{i}T^{j}]=\delta_{ij}$, then, using the representation given in Eq. \ref{gama1.fun}, the inhomogeneous BSE simplifies into a system of simultaneous integral equations describing the functions $F_{i}(p^{2},P^{2},p.P)$. To obtain masses using inhomogeneous BSEs, one straightforward approach is to focus on (Secs. 3.1 and 3.2 in Ref. \cite{GI159}).
 \begin{equation}\label{pade9.fun}
\centering
P_{E}({P^{2}})=\frac{1}{E(p^{2}=0;P^{2})}
\end{equation}
and locate its zeros. Where, $E(p^{2}=0;P^{2})$ is the BSA resulting from the numerical solution of the IBSE, as illustrated in Eq. \ref{homo.eq}.
Here, we use a diagonal Pad$\acute{e}$ approximant with its general expression for order $N$, which is presented below.
\begin{equation}\label{pade1.fun}
\centering
f_{N}(x)=\frac{c_{0}+c_{1}x+....+c_{N}x^{N}}{1+c_{N+1}x+....+c_{2N}x^{N}}
\end{equation}
It was found that \cite{GI159} a Pad$\acute{e}$ approximant can accurately identify the pole representing the ground state and the first excited states. To examine the data sample produced by Eq. \ref{pade9.fun}, defined on the discretized $P^{2}$ points, we used a Pad$\acute{e}$ approximant of order 3 to fit the sample values. The results obtained from the calculations are shown in Table ~\ref{tab:vertical_table}.\\
Similar to the PS case, we can compute the BSE for the antimeson, denoted as $\bar{\Gamma}^{du}(p;P)=[C^{-1}\Gamma^{du}(-p;P)C]^{T}$, by using the identities provided in reference \cite{GI109}. The following is the general expression for the BSA of a PS meson, such as the kaon, that is not a charge eigenstate.
\begin{equation}\label{gamau.fun}
\begin{aligned}
i \Gamma^{us}(p;P)=i\gamma_{5}E_{K}(p;P)+\gamma_{5}\gamma.PF_{K}(p;P)+\gamma_{5}\gamma.pG_{K}(p;P)+\gamma_{5}p_{\alpha}\sigma^{\alpha\beta}P_{\beta}H_{K}(p;P)
\end{aligned}
\end{equation}
 For V mesons ($J^{PC}=1^{--}$), as discussed in \cite{GI109}, we have $\bar{\Gamma}^{uu}_{\mu}(p;P) = -\Gamma^{uu}_{\mu}(p;P)$, with the negative sign resulting from the charge conjugation quantum number, C.
The general expression is represented by the following equation.
\begin{equation}\label{gamaV.fun}
\begin{aligned}
 \Gamma_{\mu}(p;P)=&\gamma_\mu V_{1}(p;P)+p_\mu \gamma \cdot p V_{2}(p;P)+p_\mu \gamma \cdot P p \cdot P V_{3}(p;P) +\gamma_5 \epsilon_{\mu \alpha \nu \beta} \gamma_\alpha p_\nu P_\beta V_{4}(p;P)\\
 &+p_\mu V_{5}(p;P)+\sigma_{\mu \nu} p_\nu p \cdot P V_{6}(p;P)+\sigma_{\mu \nu} P_\nu V_{7}(p;P)+p_\mu p_\alpha \sigma_{\alpha \beta} P_\beta V_{8}(p;P)
\end{aligned}
\end{equation}

The BSA is characterized by four independent covariant structures when dealing with an S meson with quantum numbers, $J^{PC}=0^{++}$, which is described in Ref. \cite{GI153} and can be written as:
$$
\begin{array}{ll}
T_1=1 , \hspace{0.5cm}T_2=i \gamma \cdot P, \hspace{0.5cm}T_3=i \gamma \cdot p , \hspace{0.5cm}T_4=i \sigma^{p, P}
\end{array}
$$
where $\sigma^{p, P}:=\frac{i}{2}[\gamma^{\mu} , \gamma^{\nu} ]p_{\mu}P_{\nu}$.\\ 
Once the quark propagator is known, one may continue to solve the homogeneous BSE numerically, as detailed in Ref. \cite{GI109,GI143,GI159,GI1444,GI1446}.
The homogeneous BSE represents an eigenvalue problem, and the solutions are found only for specific, discrete values of $P^{2}=-m^{2}$, and each eigenvalue corresponds to an eigenvector, which is the BSA $\Gamma_{M}^{fg}(p;P)$. 
We plot the eigenvalue spectrum as a function of $P^{2}$ and identify the meson at the point where the eigenvalue equals $1$. The results obtained from the calculations are shown in Table ~\ref{tab:table2}.\\
We have solved the inhomogeneous BSE in the rainbow truncation by employing the Pad$\acute{e}$ approximation to obtain our results. This method is advantageous because it allows us to avoid solving the equation directly in the complex plane, which is computationally intensive and can introduce numerical instability. To evaluate the accuracy of this method, we compared the results with those of homogeneous BSE. After confirming the reliability of the Pad$\acute{e}$ approximation, we will apply it to the improved BC vertex to obtain further results.\\
The RL truncation does not satisfy the full STI for the QGV, since it omits the quark-ghost kernel. Therefore, in the next Sec., we use the BC vertex, which satisfies the Abelian WTI for a trivial ghost-gluon kernel and satisfies the axWTI.
 \subsection{BSE with Ball-Chu vertex}
\begin{figure}[H]
 \centering
 \includegraphics[width=0.5\linewidth]{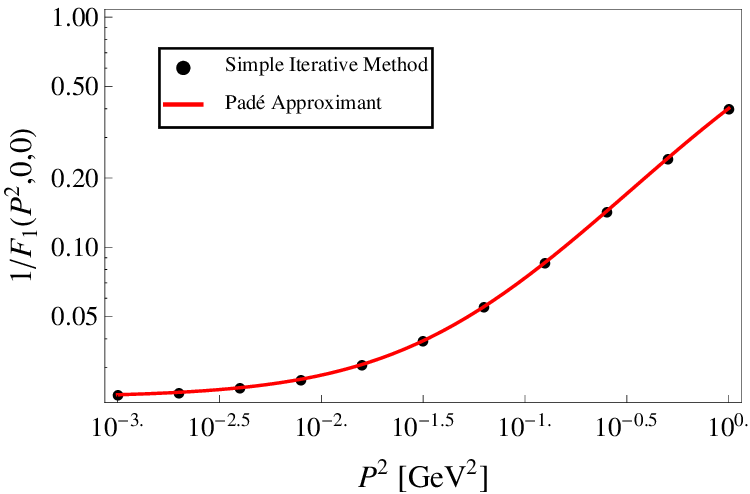}
 \caption{The value $( 1/F_1(P^2, 0, 0) )$ of the inhom. PS amplitude (Pion) computed in spacelike region using Pad$\acute{e}$ approximant vs. $P^2$.}\label{PLpsbv}
 \end{figure}

 Mesons appear as poles in the inhomogeneous BSE. The inhomogeneous BSE for the $q\bar{q}$ channel with the AV vertex, $\Gamma_{5\mu}^{fg}$ can be written as \cite{GI123,GI124,GI125,GI126,GI115,GI127,GI155}.
\begin{eqnarray}\label{bse.bc}
 \centering
\Gamma_{5\mu}^{fg}(p;P)=Z_{2}\gamma_{5}\gamma_{\mu}-g^{2}\int_{k}^{\Lambda^{2}}D_{\alpha\beta}(p-k)\frac{\lambda^{a}}{2}\gamma^{\alpha}S_{f}(k_{+})\Gamma_{5\mu}^{fg}(k;P)S_{g}(k_{-})\frac{\lambda^{a}}{2}\Gamma_{\beta}^{g}(k_{-},p_{-})
\end{eqnarray}
\begin{eqnarray*}\label{con.potential}
+g^{2}\int_{k}^{\Lambda^{2}}D_{\alpha\beta}(p-k)\frac{\lambda^{a}}{2}\gamma^{\alpha}S_{f}(k_{+})\Lambda_{5\mu\beta}^{fg}(p,k;P)
\end{eqnarray*}
Here, $\Lambda_{5\mu\beta}^{fg}$ stands for a 4-point Schwinger function. Similarly, the PS vertex, $\Gamma_{5}^{fg}(p;P),$ follows an analogous equation, and its general form is written as:
\begin{equation}\label{bas.fun}
i\Gamma^{fg}_{5}(p;P)=i\gamma_{5}E_{5}(p;P)+\gamma_{5}\gamma.PF_{5}(p;P)+\gamma_{5}\gamma.pG_{5}(p;P)+\gamma_{5}p_{\alpha}\sigma_{\alpha\beta}P_{\beta}H_{5}(p;P)
\end{equation}
To study light-quark hadrons, Eq. \ref{bse.bc} should satisfy the avWTI that is given below, where $\Gamma^{fg}_{5}(p;P)$ represents the PS vertex.
\begin{equation}
P_{\mu}\Gamma^{fg}_{5\mu}(p;P)=S_{f}^{-1}(p_{+})i\gamma_{5}+i\gamma_{5}S_{g}^{-1}(p_{-})-i[m_{f}(\zeta)+m_{g}(\zeta)]\Gamma^{fg}_{5}(p;P)
\end{equation}
This illustrates chiral symmetry and how it is broken in QCD. It has been demonstrated that the condition.
\begin{equation}\label{Pmu.fun}
P_{\mu}\Lambda^{fg}_{5\mu\beta}(p,k;P)=\Gamma_{\beta}^{f}(k_{+},p_{+})i\gamma_{5}+i\gamma_{5}\Gamma_{\beta}^{g}(k_{-},p_{-})-i[m_{f}(\zeta)+m_{g}(\zeta)]\Lambda^{fg}_{5}(p,k;P)
\end{equation}
Here, $\Lambda^{fg}_{5\beta}(p,k;P)$ is the analogue quantity to $\Lambda^{fg}_{5\mu\beta}(p,k;P)$ in the PS equation, it is both necessary and sufficient to confirm that the WTI is satisfied.
The term ``RL'' represents the leading-order component within the DSE truncation, as described in Ref. \cite{GI114}. It corresponds to $\Gamma_{\nu}^{f}(k,p) = \gamma_{\nu}$, and under this condition, Eq. \ref{Pmu.fun} is resolved with $\Lambda^{fg}_{5\mu\beta}(p,k;P) \equiv 0 \equiv \Lambda^{fg}_{5\beta}(p,k;P)$. This solution indeed represents the RL forms of the given Eq. \ref{bse.bc}.\\
In the same way that the vector WTI has been successfully employed to construct ans$\ddot{a}$tz for the fully quark-photon vertex (e.g., Ref. \cite{GI102}, \cite{GI145}), Eq. \ref{Pmu.fun} offers a means to establish a symmetry-preserving kernel for the BSE which is consistent with any reasonable Ans$\ddot{a}$tz for the fully dressed QGV in the DSE equation. The capability provided by Eq. \ref{Pmu.fun} realizes a long-standing objective in this field of study.
To illustrate, let us assume that an ans$\ddot{a}$tz for the QGV is used in Eq. \ref{Q.P} which satisfies
\begin{equation}\label{Pg.fun}
P_{\mu}i\Gamma^{f}_{\mu}(k_{+},p_{+})=\mathcal{B}(P^{2})[S_{f}^{-1}(k_{+})-S_{f}^{-1}(p_{+})]
\end{equation}
Given that $\mathcal{B}$ is flavor-independent. NB. It is important to note that, while the true QGV does not satisfy the identity given in Eq. \ref{Pg.fun} due to constraints imposed by the Slavnov-Taylor identity, which it does satisfy, a solution to Eq. \ref{Pg.fun} can still provide an acceptable pointwise estimate to the true QGV. Based on Eq. \ref{Pg.fun}, it follows that Eq. \ref{Pmu.fun}  leads to the conclusion that $(l=k-p)$,
\begin{equation}\label{PmuL.fun}
i l_{\beta} \Lambda^{fg}_{5\beta}(p,k;P)=\mathcal{B}(l^{2})[\Gamma_{5}^{fg}(k;P)-\Gamma_{5}^{fg}(p;P)]
\end{equation}
and a similar equation holds for $P_{\mu}l_{\beta}i\Lambda^{fg}_{5\mu\beta}(p,k;P)$. By solving this identity, we obtain the following equation.
\begin{equation}\label{Pg1.fun}
\Lambda^{fg}_{5\beta}(p,k;P):=-\mathcal{B}((p-k)^{2})\gamma_{5}\bar{\Lambda}^{fg}_{\beta}(p,k;P)
\end{equation}
by using Eq. \ref{bas.fun}.
\begin{equation}\label{delta.fun}
\centering
\bar{\Lambda}^{fg}_{\beta}(p,k;P)=2\ell_{\beta}[i\Delta_{E_{\pi}}(k,p;P)+\gamma.P\Delta_{F_{\pi}}(k,p;P)]+\gamma_{\beta}\bar{\Delta}_{G_{\pi}}(k,p;P)+2\ell_{\beta}\gamma.\ell\Delta_{G_{\pi}}(k,p;P)
\end{equation}
\begin{equation*}
\centering
+\frac{i}{2}[\gamma_{\beta},\gamma.P]\bar{\Delta}_{H_{\pi}}(k,p;P)+\frac{i}{2}2\ell_{\beta}[\gamma.\ell,\gamma.P]\Delta_{H_{\pi}}(k,p;P)]
\end{equation*}
If one considers an ans$\ddot{a}$tz for the QGV which satisfies Eq. \ref{Pg.fun}, the PS analogue of Eq. \ref{bse.bc}, along with Eqs. \ref{Q.P}, \ref{Pg1.fun}, and \ref{delta.fun} establish a self-contained system that preserves symmetry. Solving this system provides predictions for the properties of PS mesons \cite{GI146}.\\
 In this context, we use the BC vertex for the dressed QGV as shown in Eq. \ref{BCL}.
It states that Eq. \ref{BCL} does not impose the condition that $\mathcal{B}$ be equal to 1 in the expression given by Eq. \ref{Pg.fun}, as any change from this value can be absorbed into the fully dressed-gluon propagator.\\
To obtain masses of mesons using the inhomogeneous BSEs with the BC vertex, we employ the same method used for the inhomogeneous BSE with the bare vertex; see Fig. \ref{PLpsbv}. The results obtained from the calculations are shown in Table ~\ref{tab:vertical_table2}.
 
After determining the solution in the spacelike region, we can obtain the solution in the timelike region. In this work, we present only the ground states.
 \begin{figure}[H]
 \centering
 \includegraphics[width=0.5\linewidth]{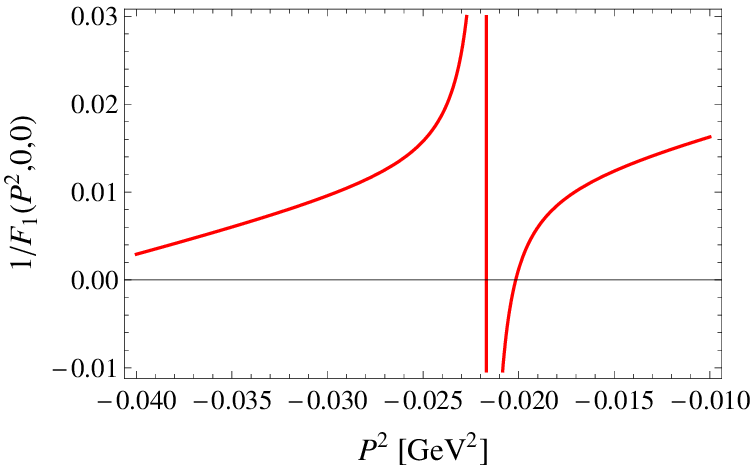}
 \caption{The value $( 1/F_1(P^2, 0, 0) )$ of the inhom. PS amplitude (Pion) computed in timelike region using Pad$\acute{e}$ approximant vs. $P^2$.}\label{PLlkplplsobv}
 \end{figure}
 Fig. \ref{PLlkplplsobv} illustrates the pole that represents a meson bound state. The position of the pole is identified by the point where the graph intersects the horizontal axis.
 \section{Results}\label{results}
The strength of the gluon models within the infrared region is set by the chiral quark condensate \cite{GI147,GI165}.
In the chiral limit, $\hat{m}= 0$ $(\mu\rightarrow \infty)$, it is represented as follows:
\begin{equation}
\centering
-\langle \bar{q}q\rangle ^{0}_{\mu}=Z_{4}N_{c}Tr\int_{k}S_{\hat{m}}(k,\mu)
\end{equation}
 By fixing the effective gluon strength, one can, in principle, obtain a more realistic result for the chiral quark condensate.
To be consistent with current estimates \cite{GI147}, we have determined the parameters of the MT gluon model, namely $\omega$ and the value of the product $\omega D$, for two different truncations: RL and BC vertex truncation. Specifically, the parameters have been adjusted so that the chiral quark condensate is $-\langle \bar{q}q\rangle ^{0}_{\mu_{19}}=(0.277$ GeV$)^{3}$ for the RL truncation and $(0.307$ GeV$)^{3}$ for the BC vertex truncation. 
This ensures that our model agrees with the experimental results. In Ref. \cite{GI125}, the ultraviolet finite model gap equation is used, whereas we used the MT model in our calculations.
\begin{table}[H]
\centering
\caption{Current-quark masses: $m_{u/d}=$3.74 MeV and $m_{s}=$825 MeV. Notes: (i) $\langle\bar{q}q\rangle^{0}$ represents the chiral quark condensate; (ii) $m_{\pi},\; m_{K},\; m_{\rho},\; m_{\sigma},\;$ are, respectively, the pion, kaon, rho and sigma meson masses obtained by solving inhomogeneous BSE. The particular selection of parameters for RL truncation are also presented. ($D$, measured in GeV$^{2}$, and other values expressed in GeV.)}
\label{tab:vertical_table}
\begin{tabular}{l|c|c|c}
\hline \hline
{Quantity} & {This Work} & {Ref. \cite{GI125}} & {Exp} \\
\hline
Vertex & RL & RL & - \\
$\omega$ & 0.4 & 0.5 & - \\
$D$ & 0.93 & 0.5 & - \\
$-(\langle \bar{q}q\rangle^{0})^{1/3}$ & 0.277 & 0.13 & - \\
$m_{\pi}$ & 0.139 & 0.160 & 0.138 \\
$m_{K}$ & 0.501 & - & 0.496 \\
$m_{\rho}$ & 0.814 & - & 0.770 \\
$m_{\sigma}$ & 0.670 & 0.27 & 0.4-1.2 \\
\hline \hline
\end{tabular}
\end{table}
\begin{table}[H]
 \begin{center}
 \caption{Masses obtained by solving homogeneous BSE using ladder truncation.}
 \label{tab:table2}
\begin{tabular}{c|c|c|c|c|c|c|c}
\hline \hline Vertex  & $\omega$ & $D$ & $-(\langle \bar{q}q\rangle^{0}) ^{1/3}$ & $m_{\pi}$ & $m_{K}$ & $m_{\rho}$& $m_{\sigma}$ \\
RL& 0.4 &0.93& 0.277& 0.138& 0.494 &0.802 &0.668\\
\hline \hline
 \end{tabular}
\end{center}

\end{table}
\begin{table}[H]
\centering
\caption{Masses obtained by solving inhomogeneous BSE using BC vertex. The parameters for the BC vertex are also presented.}
\label{tab:vertical_table2}
\begin{tabular}{l|c|c|c}
\hline \hline
Quantity & This Work & Ref. \cite{GI125} & Exp \\
\hline
Vertex & BC & BC & -\\
$\omega$ & 0.4 & 0.5 & -\\
$D$ & 0.406 & 0.5 & -\\
$-(\langle \bar{q}q\rangle^{0})^{1/3}$ & 0.307 & 0.26 &- \\
$m_{\pi}$ & 0.140 & 0.14 & 0.138 \\
$m_{K}$ & 0.474 & - & 0.496 \\
\hline \hline
\end{tabular}
\end{table}
\section{Conclusions}
This paper focuses on the meson spectrum, studied through the Bethe-Salpeter and Dyson-Schwinger equations. DSE of the quark propagator is truncated using the MT model for the dressed gluon propagator, which incorporates both the infrared enhancement and perturbatively correct ultraviolet behavior. Additionally, for the dressed quark-gluon vertex, we apply its bare form and the BC model, which minimally satisfies the constraint imposed by the gauge symmetry through its Ward-Green-Takahashi identity. Consistent truncation of both Bethe-Salpeter and Dyson-Schwinger equations requires that avWTI, arising from chiral symmetry of the Lagrangian of QCD, must not be violated. We utilize this identity as an additional constraint to truncate BSE and examine the impact of two choices of QGV, along with the application of the MT model on the light meson spectrum.
Advanced matrix algorithms were used to obtain the numerical solution of both the homogeneous and inhomogeneous vertex BSE. The homogeneous BSE represents an eigenvalue problem; we plot the eigenvalue spectrum as a function of $p^2$ and identify the meson at the point where the eigenvalue equals 1. The inhomogeneous BSE was considered as an inhomogeneous linear system, with the bound state mass appears as a pole in the inhomogeneous BSE. The pole position was calculated through a fit of the Pad$\acute{e}$ approximation to the BSA, which enables us to locate the poles of the Bethe-Salpeter amplitude without requiring a numerical solution in the timelike region of momentum space. This approach enabled the identification of both the ground state pole and additional poles representing various excitations. However, the present investigation focused exclusively on the ground state pole.\\ 
 In case of RL truncation, we applied the framework to compute the masses of PS, S, and V mesons. In the case of the BC vertex, we studied the PS meson spectrum. Specifically, we examine the flavored and unflavored mesons. 
We find that consistent truncations of DSE and BSE based on bare and BC vertices, coupled with the MT model, yield consistent results for both flavored and unflavored PS mesons. 
Our next course of action involves employing the BC vertex to determine the masses of S and V mesons.
 An imminent objective is to formulate a reliable Bethe-Salpeter kernel for a fully-dressed QGV describing S and V mesons, which we consider to be the forthcoming challenge. Consequently, we strongly believe that truncations beyond RL have considerable significance for numerous hadron characteristics. 
 This covers features such as the excited state spectrum and the electromagnetic elastic and transition form factors of the nucleon, as shown in references \cite{GI163,GI164}. Some of these aspects are currently being studied and will be addressed in future reports.
\begin{acknowledgments}
Authors acknowledge University of the Punjab, Lahore
for providing a supportive environment. F.A. acknowledge financial support of HEC under grant 20-15728/NRPU/R\&D/HEC/2021.  
\end{acknowledgments}

\begin{appendix}

\end{appendix}

\end{document}